\documentclass[prc,twocolumn,floatfix,groupedaddress,nofootinbib,showpacs,preprintnumbers,
amsmath,amssymb,amsfonts,superscriptaddress,widetable] {revtex4-1}
\usepackage{bm}
\usepackage{mathrsfs}
\usepackage{graphicx}
\usepackage{array}
\usepackage{xcolor}
\usepackage{relsize}

\newcommand{\Rskin}[1]{$R_{\rm skin}^{#1}$}
\newcommand{\AlphaD}[1]{$\alpha_{\raisebox{1pt}{\tiny D}}^{#1}$}
\begin{document}
\title{Implications of PREX-2 on the electric dipole polarizability 
        of neutron rich nuclei}
\author{J. Piekarewicz}\email{jpiekarewicz@fsu.edu}
\affiliation{Department of Physics, Florida State University,
               Tallahassee, FL 32306, USA}
\date{\today}
\date{\today}
\begin{abstract}

\noindent\textbf{Background:} The recent announcement by the PREX collaboration 
of an unanticipated thick neutron skin in ${}^{208}$Pb (\Rskin{208}) has challenged 
our understanding of neutron rich matter in the vicinity of nuclear saturation density. 
Whereas earlier constraints indicate that the symmetry energy is relatively soft, the 
PREX-2 result seems to suggest the opposite. 
\smallskip

\noindent\textbf{Purpose:} To confront constraints on the symmetry energy obtained
from measurements of the electric dipole polarizability against those informed by the 
PREX-2 measurement of \Rskin{208} and by the correlations that it entails.
\smallskip

\noindent\textbf{Methods:} Covariant energy density functionals informed by the 
properties of finite nuclei are used to compute the electric dipole response of 
${}^{48}$Ca, ${}^{68}$Ni, ${}^{132}$Sn, and ${}^{208}$Pb. The set of functionals
used in this work are consistent with experimental data, yet are flexible enough in 
that they span a wide range of values of \Rskin{208}. 
\smallskip

\noindent\textbf{Results:} It is found that theoretical predictions of the electric dipole 
polarizability that are consistent with the PREX-2 measurement systematically 
overestimate the corresponding values extracted from the direct measurements of the
distribution of electric dipole strength. 
\smallskip

\noindent\textbf{Conclusions:} The neutron skin thickness of ${}^{208}$Pb extracted
from parity violating electron scattering and the electric dipole polarizability measured
in photoabsorption experiments are two of the cleanest experimental tools used to constrain
the symmetry energy around nuclear saturation density. However, the recent value of 
\Rskin{208} that suggests a fairly stiff symmetry energy stands in stark contrast to the 
conclusions derived from the electric dipole polarizability. At present, we offer no solution 
to this dilemma.
\end{abstract}

\smallskip

\maketitle

\section{Introduction}
\label{Sec:Introduction}

Experiments with electroweak probes provide an effective strategy to elucidate the 
complex nuclear dynamics. This stands in contrast to experiments involving hadronic 
probes that are often plagued by considerable model dependencies and uncontrolled 
approximations\,\cite{Thiel:2019tkm}. 
Beside their intrinsic value 
in the determination of fundamental nuclear properties, electroweak experiments provide 
a clean and powerful link to the equation of state (EOS) of neutron rich matter. Indeed, 
laboratory experiments provide the first rung in a ``density ladder" that aims to determine 
the EOS over a wide range of densities.

The quest to determine the EOS has been reenergized by the plethora of recent astronomical 
discoveries that include the identification of massive neutron 
stars\,\cite{Demorest:2010bx,Antoniadis:2013pzd,Cromartie:2019kug,Fonseca:2021wxt}, 
the simultaneous determination of the mass and radius of two neutron 
stars\,\cite{Riley:2019yda,Miller:2019cac,Miller:2021qha,Riley:2021pdl,Raaijmakers:2021uju}, 
and the detection of gravitational waves from a binary neutron star merger\,\cite{Abbott:PRL2017}. 
These historic discoveries have placed stringent constraints on the EOS at densities above 
two-to-three times nuclear saturation density. 

Among the electroweak experiments that have been identified as having a strong impact on the EOS
of neutron-rich matter (i.e., strong isovector indicators) are parity-violating electron 
scattering\,\cite{Abrahamyan:2012gp,Horowitz:2012tj,Adhikari:2021phr} and photoabsorption 
reaction\,\cite{Tamii:2011pv,Tamii:2013cna,Wieland:2009,Rossi:2013xha,Hashimoto:2015ema,
Birkhan:2016qkr,Tonchev:2017ily,Bassauer:2020iwp}. For such class of experiments, the connection 
to the equation of state emerges from a correlation between the slope of the symmetry energy at 
saturation density ($L$) and both the neutron skin thickness of $^{208}$Pb (\Rskin{208})\,\cite{Brown:2000,
Furnstahl:2001un, Centelles:2008vu,RocaMaza:2011pm} and the product of the electric dipole 
polarizability \AlphaD{} times the value of the nuclear symmetry energy at saturation density 
$J$\,\cite{Satula:2005hy,Roca-Maza:2013mla,Roca-Maza:2015eza}. 

Given the importance of the symmetry energy in constraining a host of neutron-star 
properties---particularly stellar radii\,\cite{Lattimer:2006xb}---a concerted community effort 
has been devoted to determine the symmetry energy and its slope at saturation density. For 
example, from one of the latest compilations\,\cite{Drischler:2020hwi}, the recommended 
values for these two parameters are
\begin{subequations}
\begin{align}
  & J=(31.7\pm1.1)\,{\rm MeV}, \\
  & L=(59.8\pm4.1)\,{\rm MeV}.
\end{align}
\label{JLDrischler} 
\end{subequations}
The above values are in good agreement with other estimates that were obtained from either 
purely theoretical approaches or extracted from a theoretical interpretation of experimental 
data\,\cite{Chen:2010qx,Steiner:2011ft,Lattimer:2012nd,Zhang:2013wna,Hebeler:2013nza,
Gandolfi:2013baa,Hagen:2015yea,Roca-Maza:2015eza,Drischler:2020hwi}; see
Fig.\,2 of Ref.\,\cite{Drischler:2020hwi}. Based on such a preponderance of evidence, it was 
concluded that the symmetry energy in the vicinity of nuclear saturation density is relatively 
soft. That is, the pressure increases slowly with increasing density.

However, all this evidence was collected prior to the completion of the latest Lead Radius 
EXperiment (PREX-2) at Jefferson Lab\,\cite{Adhikari:2021phr}. Due to problems during 
the first phase of the experiment (``PREX-1") the reported error bars at that time were simply too 
large to place any meaningful constraint on the equation of state. Indeed, the first report from the 
PREX collaboration\,\cite{Abrahamyan:2012gp}---together with the strong linear correlation 
between \Rskin{208} and $L$ identified in Ref.\,\cite{RocaMaza:2011pm}---yielded the 
following results:
\begin{subequations}
\begin{align}
  &\text{\Rskin{208}} = \Big(\!0.33^{+0.16}_{-0.18}\Big)\,{\rm fm}, \\
  & \hspace{8pt} 35 \lesssim\!L({\rm MeV})\!\lesssim 265.
\end{align}
\end{subequations}

The PREX-2 campaign has dramatically changed the landscape by delivering on the original promise 
to determine the neutron radius of ${}^{208}$Pb with a $\sim\!0.06\,{\rm fm}$ (or 1\%) precision. By 
combining the PREX-1 and PREX-2 measurements, the improved value of  \Rskin{208} was reported 
to be\,\cite{Adhikari:2021phr}
\begin{equation}
 R_{\rm skin}=(0.283\pm0.071)\,{\rm fm}.
 \label{Rskin}
\end {equation} 
Using this newly reported value---together with the strong correlation between \Rskin{208} and 
both the symmetry energy $J$ and its slope $L$ at saturation density---the following  1$\sigma$ 
intervals were obtained\,\cite{Reed:2021nqk}:
\begin{subequations}
\begin{align}
 & J = (38.1 \pm 4.7) {\rm MeV}, \label{RMFJ}\\
 & L = (106 \pm 37) {\rm MeV} \label{RMFL}.
\end{align} 
\label{JandL}
\end{subequations}
These updated results greatly overestimate the central value and uncertainty quoted in 
Eq.(\ref{JLDrischler}). Given that some of the earlier limits on both $J$ and $L$ were inferred 
from the analysis of the electric dipole polarizability of a few neutron-rich nuclei, it is the aim of 
this paper to confront those earlier results against the new limits obtained from incorporating 
the PREX-2 recommended value for \Rskin{208}. 

The manuscript has been organized as follows. In Sec.\ref{Sec:Formalism} we review some 
of the most salient features of the theoretical framework used to compute the electric dipole 
response. Predictions, informed by the recent PREX-2 measurement, are made in 
Sec.\ref{Sec:Results} for the electric dipole 
polarizability of ${}^{48}$Ca, ${}^{68}$Ni, ${}^{132}$Sn, and ${}^{208}$Pb. These predictions
are later compared against the experimentally recommended values obtained from photoabsorption 
experiments. Finally, we present a short summary and conclusions in Sec.\ref{Sec:Conclusions}.
\vfill

\section{Formalism}
\label{Sec:Formalism}

The theoretical framework implemented in this work has been presented in much greater detail 
in several references, so we limit ourselves to highlight the most relevant points; see for 
example\,\cite{Piekarewicz:2013bea,Yang:2019fvs} and references contained therein. The starting 
point for the relativistic calculation of the nuclear response is the covariant model of Ref.\,\cite{Mueller:1996pm} 
supplemented by an isoscalar-isovector term originally introduced in Ref.\,\cite{Horowitz:2000xj}. 
That is, the interacting Lagrangian density is given by
\begin{widetext}
\begin{eqnarray}
 {\mathscr L}_{\rm int} &=&
\bar\psi \left[g_{\rm s}\phi   \!-\! 
         \left(g_{\rm v}V_\mu  \!+\!
    \frac{g_{\rho}}{2}{\mbox{\boldmath $\tau$}}\cdot{\bf b}_{\mu} 
                               \!+\!    
    \frac{e}{2}(1\!+\!\tau_{3})A_{\mu}\right)\gamma^{\mu}
         \right]\psi \nonumber \\
                   &-& 
    \frac{\kappa}{3!} (g_{\rm s}\phi)^3 \!-\!
    \frac{\lambda}{4!}(g_{\rm s}\phi)^4 \!+\!
    \frac{\zeta}{4!}   g_{\rm v}^4(V_{\mu}V^\mu)^2 +
   \Lambda_{\rm v}\Big(g_{\rho}^{2}\,{\bf b}_{\mu}\cdot{\bf b}^{\mu}\Big)
                           \Big(g_{\rm v}^{2}V_{\nu}V^{\nu}\Big)\;.
 \label{LDensity}
\end{eqnarray}
\end{widetext}
The Lagrangian density includes as the basic degrees of freedom an isodoublet 
nucleon field $\psi$ interacting through the exchange of photons ($A_{\mu}$) and 
three massive ``mesons": a scalar-isoscalar ($\phi$) a vector-isoscalar ($V^{\mu}$), 
and a vector-isovector (${\bf b}_{\mu}$)\,\cite{Walecka:1974qa,Serot:1984ey,Serot:1997xg}.
The isoscalar-isovector coupling constant $\Lambda_{\rm v}$ is highly sensitive to 
the density dependence of symmetry energy---and in particular to its slope at 
saturation density $L$\,\cite{Horowitz:2000xj}. Useful to the calibration of the
model is the existence of an analytic---one-to-one---correspondence between 
bulk parameters of infinite nuclear matter and the various coupling constants. In 
particular, there is a one-to-one correspondence between the two isovector 
parameters ($g_{\rho}$ and $\Lambda_{\rm v}$) and the value of symmetry 
energy $J$ and its slope $L$ at saturation density\,\cite{Chen:2014sca}.

Once the calibration of the functional is completed, one proceeds to compute
ground-state properties of finite nuclei by solving self-consistently the resulting 
Kohn-Sham equations\,\cite{Yang:2019fvs}. Among the ground state properties 
that emerge is a set of single-particle energies, a corresponding set of Dirac orbitals, 
the Kohn-Sham potentials, and neutron and proton densities. 
In turn, the dynamic response of the system to an external probe is computed in 
a relativistic random phase approximation (RPA). The consistent nuclear response
of the ground state is encoded in the RPA formalism that ensures that important 
symmetries are maintained\,\cite{Piekarewicz:2013bea}. Indeed, by employing a 
residual particle-hole interaction consistent with the interaction used to generate the 
mean-field ground state, the conservation of the vector current is preserved
and the spurious strength associated with a uniform translation of the center
of mass is decoupled from the physical response\,\cite{Piekarewicz:2013bea}. 

The cornerstone of the theoretical framework is the polarization tensor, whose 
imaginary part is directly connected to the experimentally accessible nuclear 
response\,\cite{Fetter:1971,Dickhoff:2005}. In particular, the distribution of electric 
dipole ($E1$) strength of interest in this work can be isolated from the longitudinal 
response that is defined as follows:
\begin{align}
  S_{L}(q,\omega)\!=\!\sum_{n}\Big|\langle\Psi_{n}|\hat{\rho}({\bf q})|
  \Psi_{0}\rangle\Big|^{2}\!\delta(\omega\!-\!\omega_{n}).
\label{SLong}
\end{align}
Here $\Psi_{0}$ is the exact nuclear ground state, $\Psi_{n}$ is an excited state 
with excitation energy $\omega_{n}\!=\!E_{n}\!-\!E_{0}$, and $\hat{\rho}({\bf q})$ is 
the transition operator that is given as the Fourier transform of the vector-isovector 
density. That is, 
\begin{equation}
  \hat{\rho}_{a}({\bf q}) \!=\! \int d^{3}r \, \bar{\psi}({\bf r}) 
  e^{-i{\bf q}\cdot{\bf r}} \gamma^{0} \tau_{3}\psi({\bf r}),
 \label{Rhoq}
\end{equation}
where $\gamma^{0}\!=\!{\rm diag}(1,1,-1,-1)$ is the zeroth component of the Dirac 
matrices and $\tau_{3}\!=\!{\rm diag}(1,-1)$ is the third component of the Pauli
matrices in isospin space. 

To lowest order, the uncorrelated polarization describes particle-hole excitations 
between the single-particle orbitals obtained from solving the Kohn-Sham equations. 
The RPA response goes beyond the uncorrelated response by building collectivity 
through the mixing of all particle-hole excitations with the same quantum numbers. 
If many particle-hole pairs close in energy are involved and the residual particle-hole 
interaction is strong, then the nuclear response is strongly collective and one 
``giant resonance" dominates. In the non-spectral framework employed in this work, 
the continuum width of the resonance---obtained from exciting a bound nucleon into 
the continuum---is treated exactly. Not so, however, the spreading width, which is 
driven by more complex configurations. This is a well-known deficiency of the RPA 
approach.

In the long wavelength limit, the distribution of isovector dipole strength $R(\omega;E1)$
is related to the longitudinal response through the following equation:
\begin{equation}
 R(\omega;E1) = 
 \lim_{q\rightarrow 0} \left(\frac{9}{4\pi q^{2}}\right)S_{L}(q,\omega;E1)\;,
 \label{RGDR} 
\end{equation}
which in turn, is directly related to the photoabsorption cross section 
\begin{equation}
 \sigma_{\!\rm abs}(\omega) = \frac{16\pi^{3}}{9}\frac{e^{2}}{\hbar c}
 \omega R(\omega;E1).
\label{PhotoAbs}
\end{equation}
Widely used in the literature are the following moments of the distribution of strength: 
(a) the energy weighted sum $m_{1}$; (b) the energy unweighted sum $m_{0}$, and (c) 
the inverse energy weighted sum $m_{-1}$, which is proportional to the electric dipole 
polarizability [see Eq.(\ref{alphaD})]. Centroid energies of collective resonances are defined 
as $E_{\rm c}\!=\!m_{1}/m_{0}$, with the $m_{1}$ moment ``protected" by the (largely) 
model-independent energy weighted sum rule (EWSR)\,\cite{Harakeh:2001}:
\begin{equation}
 m_{1} =
 \frac{9\hbar^{2}}{8\pi M}\left(\frac{NZ}{A}\right)\!\approx\!
 14.8 \left(\frac{NZ}{A}\right) {\rm fm}^{2}{\rm MeV} \;.
 \label{EWSR}
\end{equation}
Being model independent, the EWSR is not effective at discriminating among various 
theoretical models. Instead, the electric dipole polarizability has been identified as a strong 
isovector indicator\,\cite{Reinhard:2010wz} that is highly sensitive to the stiffness of the
symmetry energy\,\cite{Piekarewicz:2012pp}. The electric dipole polarizability is defined as
follows:
\begin{equation}
 \text{\AlphaD{}}  = \frac{\hbar c}{2\pi^{2}} \int_{0}^{\infty} 
 \frac{\sigma_{\!\rm abs}(\omega)}{\omega^{2}}\,d\omega =
 \frac{8\pi e^2}{9} m_{-1}. 
\label{alphaD}
\end{equation}
It is the powerful connection between the photoabsorption cross section $\sigma_{\!\rm abs}(\omega)$,
\AlphaD{}, and $L$ that we will examine in the next section. 

\section{Results}
\label{Sec:Results}

In this section we present results for the electric dipole response of four neutron-rich nuclei:
${}^{48}$Ca, ${}^{68}$Ni, ${}^{132}$Sn, and ${}^{208}$Pb. Given that our main goal is to
constrain bulk properties of the symmetry energy, these four spherical nuclei were chosen to 
minimize uncertainties associated with certain nuclear-structure effects, such as pairing 
correlations. Moreover, with the exception of ${}^{132}$Sn, data already exists for the electric 
dipole polarizability of ${}^{208}$Pb\,\cite{Tamii:2011pv}, ${}^{68}$Ni\,\cite{Wieland:2009,Rossi:2013xha},
and ${}^{48}$Ca\,\cite{Birkhan:2016qkr}. 

To assess the impact on the electric dipole polarizability from the combined PREX-1--PREX-2 
measurements (henceforth referred simply as ``PREX-2")  we make predictions for the distribution 
of electric dipole strength using a set of covariant energy density functionals (EDFs) that span a 
wide range of values for the neutron skin thickness of ${}^{208}$Pb. These functionals include 
FSUGold2\,\cite{Chen:2014sca}, together with a set of six systematically varied 
interactions---FSUGold2--L047, L054, L058, L076, L090, L100---that have identical isoscalar 
properties as FSUGold2, but isovector properties defined by a preselected value for the slope of 
the symmetry energy\,\cite{Reed:2021nqk}. For example, FSUGold2--L100 was calibrated with
an input value of $L\!=\!100\,{\rm MeV}$. Another set of accurately calibrated density functionals 
is given by RMF016, RMF022, RMF032\,\cite{Chen:2014mza}, where now the labels are associated 
to the predicted value of \Rskin{208}. For example, RMF032 predicts a neutron skin thickness of 
\Rskin{208}$\!=\!0.32\,{\rm fm}$. Finally, TFa, TFb, and TFc, with \Rskin{208}$\!=\!0.25, 0.30, 
{\rm and}\, 0.33\,{\rm fm}$, respectively, were calibrated to test whether the large central value of 
\Rskin{208}$\!=\!0.33\,{\rm fm}$ originally reported by the PREX collaboration\,\cite{Abrahamyan:2012gp} 
was incompatible with other laboratory experiments and/or astrophysical observations\,\cite{Fattoyev:2013yaa}. 
Now as then, we find no compelling reason to rule out models with large neutron skins.
Additional details on the fitting protocol---including a description of the model and the 
observables used in the calibration procedure---may be found in 
Refs.\,\cite{Chen:2014sca,Chen:2014mza,Fattoyev:2013yaa}. 

The models considered here span a wide range of values for both the neutron skin thickness of 
${}^{208}$Pb and the associated slope of the symmetry energy $L$. Such a range is comparable 
to the one used in the multi-model analysis of the sensitivity of the symmetry energy to the electric 
dipole polarizability and weak-charge form factor of both ${}^{48}$Ca and 
${}^{208}$Pb\,\cite{Piekarewicz:2012pp,Reinhard:2013fpa}. Nevertheless, given that the 
present study relies on a specific class of covariant EDFs, all the results presented here 
should be tested against other theoretical approaches.

\subsection{Neutron Skins: Data-to-Data Relations}

\begin{figure}[ht]
 \centering
 \includegraphics[width=0.4\textwidth]{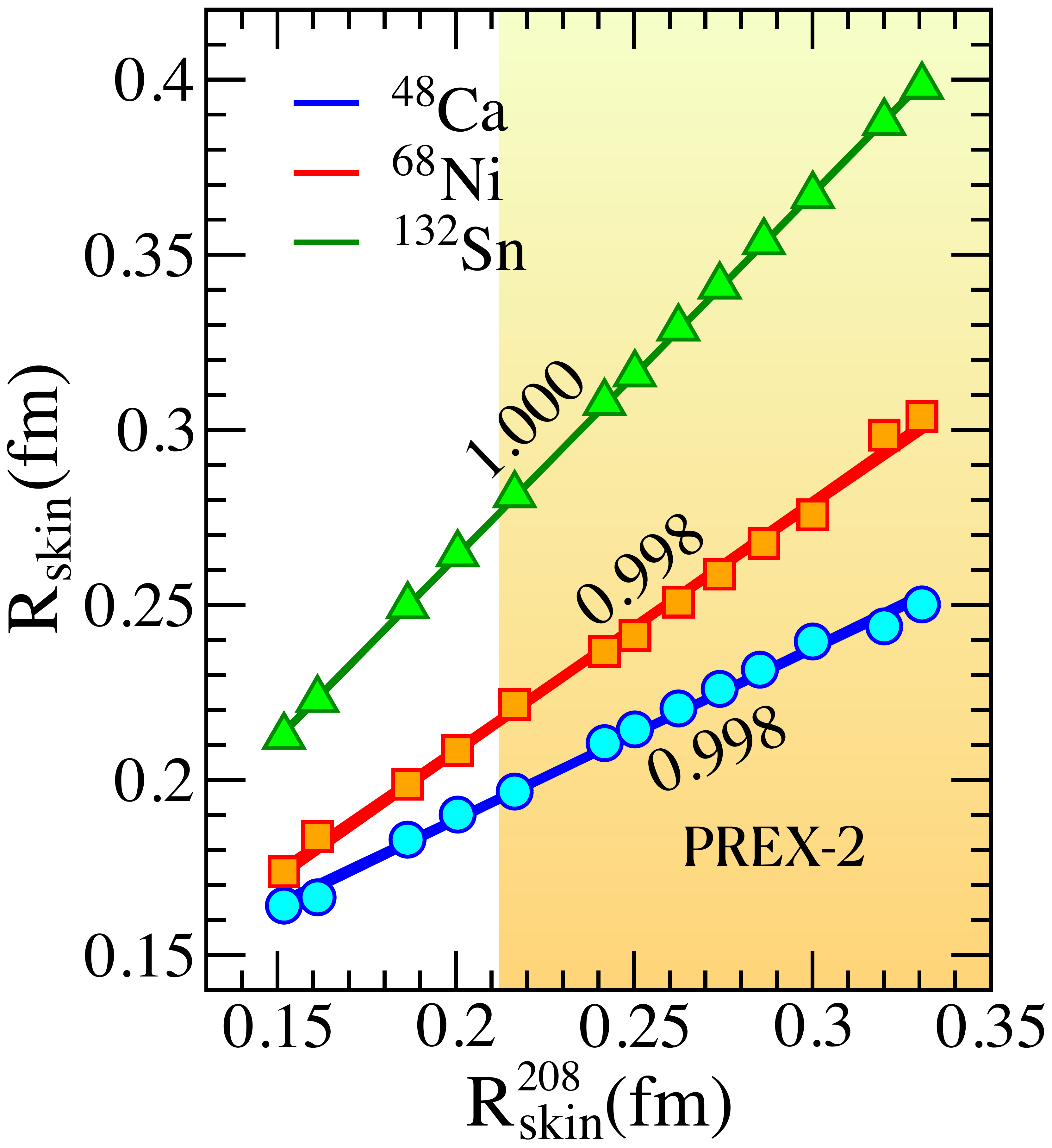}
 \caption{(Color online). Emergence of a \emph{data-to-data} relation between the neutron skin thickness of 
  ${}^{208}$Pb and the corresponding neutron skin thickness of ${}^{48}$Ca, ${}^{68}$Ni, and ${}^{132}$Sn. 
  The numbers above the lines indicate the value of the correlation coefficients and the shaded area denotes 
  the 1$\sigma$ interval reported by PREX-2\,\cite{Adhikari:2021phr}.}
\label{Fig1}
\end{figure}

Although the primary goal of this work is to confront constraints on the density dependence of the 
symmetry energy extracted from measurements of the electric dipole polarizability against those 
derived from PREX-2, we start by displaying in Fig.\ref{Fig1} predictions for the neutron skin thickness 
of ${}^{48}$Ca, ${}^{68}$Ni, and ${}^{132}$Sn inferred from the PREX-2 measurement. The numbers 
associated with each straight line represent the correlation coefficient (nearly one in all three cases) 
and the shaded region indicates the 1$\sigma$ PREX-2 error in \Rskin{208}\,\cite{Adhikari:2021phr}. 
Relying on the nearly perfect correlations displayed in the figure, the following ``data-to-data" relations 
are inferred for the neutron skin thickness of these three neutron rich nuclei:
\begin{subequations}
\begin{align}
 & \text{\Rskin{\,48}}\!=\!(0.229\pm0.035)\,{\rm fm}, \label{Rskin48} \\
 & \text{\Rskin{\,68}}\!=\!(0.267\pm0.050)\,{\rm fm}, \label{Rskin68}\\
 & \text{\Rskin{132}}\!=\!(0.350\pm0.074)\,{\rm fm} \label{Rskin132}. 
\end{align}
\end{subequations}
The prediction for ${}^{48}$Ca is particularly timely given that the \emph{Calcium Radius EXperiment} 
(CREX) has been completed and the analysis is currently under way. CREX measured the weak-charge 
form factor of ${}^{48}$Ca with a high-enough precision to allow for the extraction of the neutron radius with 
an anticipated experimental error of $\sim\!0.03$\,fm\,\cite{CREX:2013}, that is similar to the one
quoted in Eq.(\ref{Rskin48}). 

Part of the appeal of ${}^{48}$Ca---and a main motivation behind the experiment\,\cite{CREX:2013}---is that it 
provides a powerful bridge between microscopic approaches and density functional theory. In the particular 
case of ab-initio approaches rooted in coupled cluster theory, the prediction of  
$0.12\,{\rm fm}\!\lesssim\text{\Rskin{48}}\!\lesssim\!0.15\,{\rm fm}$\,\cite{Hagen:2015yea}, 
lies well outside the 1$\sigma$ region quoted in Eq.(\ref{Rskin48}). On the other hand, a combined
theoretical-experimental approach based on a dispersion-optical model predicts a best-fit value that is 
significantly larger:  $\text{\Rskin{48}}\!=\!(0.249\pm0.023)\,{\rm fm}$\,\cite{Mahzoon:2017fsg}, a value
that was later revised to $\text{\Rskin{48}}\!=\!(0.22\pm0.03)\,{\rm fm}$\,\cite{Pruitt:2020zbf}. 
The revised value is in excellent agreement with the PREX-2 informed value quoted in Eq.(\ref{Rskin48}). 

We note that the determination of the neutron skin thickness of a variety of neutron-rich nuclei is one of 
the main science drivers of the Facility for Rare Isotope Beams (FRIB)\,\cite{LongRangePlan}. Given that 
many of these experiments will involve unstable nuclei in inverse kinematics, CREX and PREX-2 will provide 
critical anchors for the calibration of hadronic reactions.

\subsection{Correlating \Rskin{208} to $J$\AlphaD{}}

\begin{figure}[ht]
 \centering
 \includegraphics[width=0.45\textwidth]{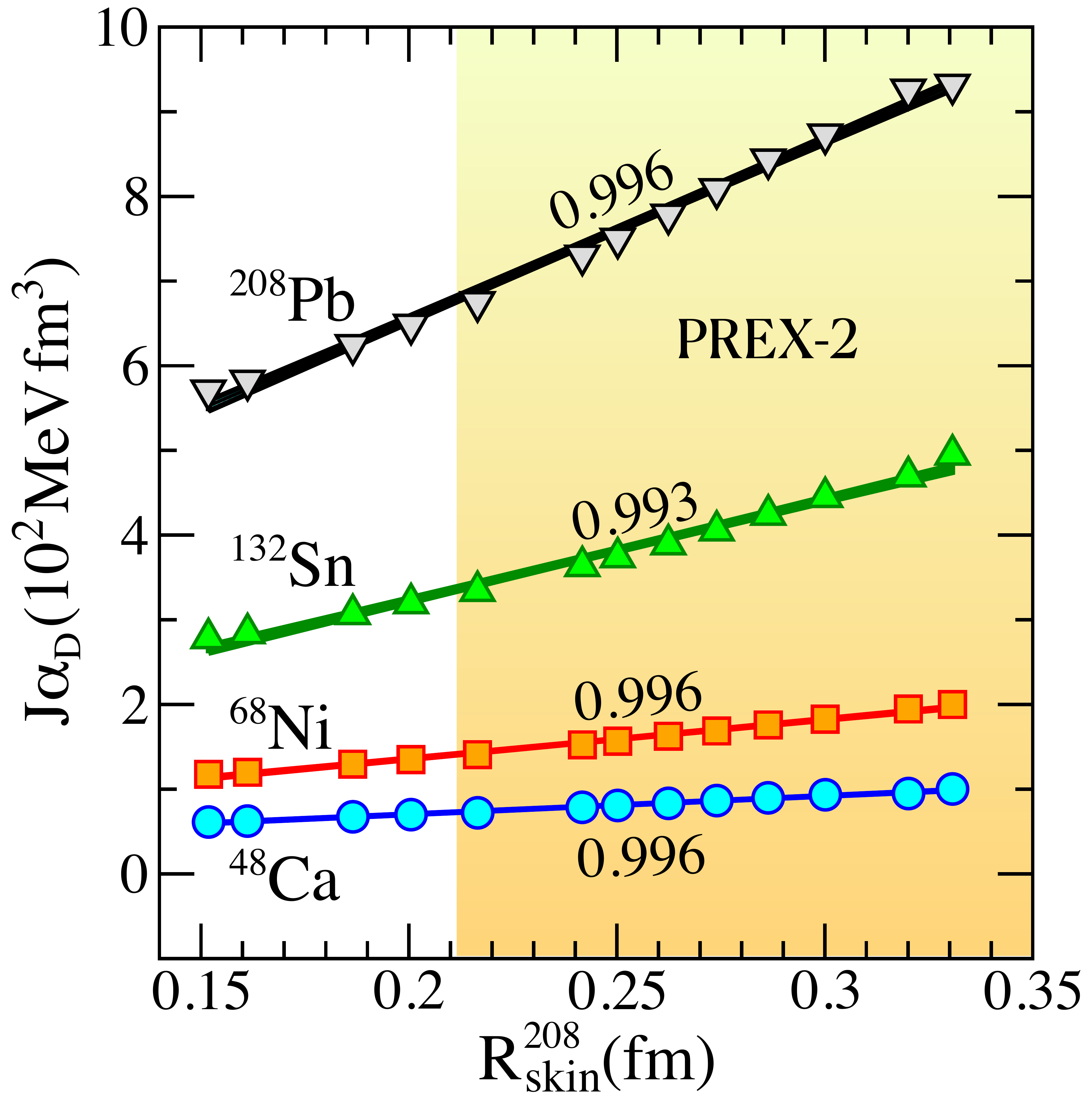}
  \caption{(Color online). Correlation between the neutron skin thickness of ${}^{208}$Pb and the product
  of the nuclear symmetry energy at saturation density times the electric dipole polarizability of ${}^{48}$Ca, 
  ${}^{68}$Ni, ${}^{132}$Sn, and ${}^{208}$Pb. The numbers above the lines indicate the value of the correlation 
  coefficients and the shaded area denotes the 1$\sigma$ error reported by PREX-2\,\cite{Adhikari:2021phr}.}
\label{Fig2}
\end{figure}

The electric dipole polarizability was identified by Reinhard and Nazarewicz as a strong isovector 
indicator that is highly sensitive to the density dependence of the symmetry energy\,\cite{Reinhard:2010wz}. 
Whereas the correlation between the electric dipole polarizability and the neutron skin thickness
is strong, a far stronger correlation exists between \Rskin{208} (or $L$) and the product of 
$J$ times the electric dipole polarizability\,\cite{Satula:2005hy,Roca-Maza:2013mla}. Indeed,
Fig.\,\ref{Fig2} displays the nearly perfect correlation between \Rskin{208} and $J$\AlphaD{} for 
all four nuclei considered in this work. As in Fig.\ref{Fig1}, the numbers next to the lines display the 
correlation coefficients and the shaded region indicates the PREX-2 error\,\cite{Adhikari:2021phr}.  
Relying on the nearly perfect correlation between  \Rskin{208} and $J$\AlphaD{}, we can extract the
following 1$\sigma$ intervals for all four nuclei of interest:
\begin{subequations}
\begin{align}
 & J\text{\AlphaD{48}}\!=\!(88.4\pm15.4)\,{\rm MeV\,fm^{3}}, \label{JAlphaD48} \\
 & J\text{\AlphaD{68}}\!=\!(174\pm33.0)\,{\rm MeV\,fm^{3}}, \label{JAlphaD68} \\
 & J\text{\AlphaD{132}}\!=\!(422\pm84.6)\,{\rm MeV\,fm^{3}}, \label{JAlphaD132} \\
 & J\text{\AlphaD{208}}\!=\!(830\pm150)\,{\rm MeV\,fm^{3}}. \label{JAlphaD208} 
\end{align}
\end{subequations}
These numbers will become important for the statstical analysis that will be carried out in the following sections.

\subsection{Electric Dipole Response of ${}^{208}$Pb}

In this section we explore the impact of PREX-2 on the electric dipole polarizability of ${}^{208}$Pb. 
A high-precision measurement of the electric dipole response of ${}^{208}$Pb was obtained from the 
small-angle $({\bf p},{\bf p}')$ scattering experiment carried out at the RCNP facility in Osaka, 
Japan\,\cite{Tamii:2011pv}. Using two independent techniques---plus a careful comparison against 
existing photoabsorption data---the following value for the electric dipole polarizability was obtained:
$\text{\AlphaD{208}}\!=\!(20.1\pm0.6)\,{\rm fm}^{3}$.
The precision attained by the RCNP experiment, together with the strong correlation between the 
neutron skin thickness and the electric dipole polarizability identified in Ref.\,\cite{Reinhard:2012vw}, 
resulted in the following recommended value for the neutron skin thickness of ${}^{208}$Pb\,\cite{Tamii:2011pv}: 
\begin{equation}
 \text{\Rskin{208}} = 0.156^{+0.025}_{-0.021}\,{\rm fm}.
 \label{RskinRCNP}
\end {equation} 
Following the same approach described in Ref.\,\cite{Reinhard:2012vw}---but now using a large and 
diverse set of EDFs---a slightly revised value was obtained:
$0.13\!\leq\!\text{\Rskin{208}}\!\leq\!0.19\,{\rm fm}$\,\cite{Roca-Maza:2015eza}, confirming the initial
estimate deduced from the RCNP experiment. These values, however, stand in stark contrast to the
much larger value reported by the PREX collaboration\,\cite{Adhikari:2021phr}.

\begin{figure}[ht]
 \centering
 \includegraphics[width=0.49\textwidth]{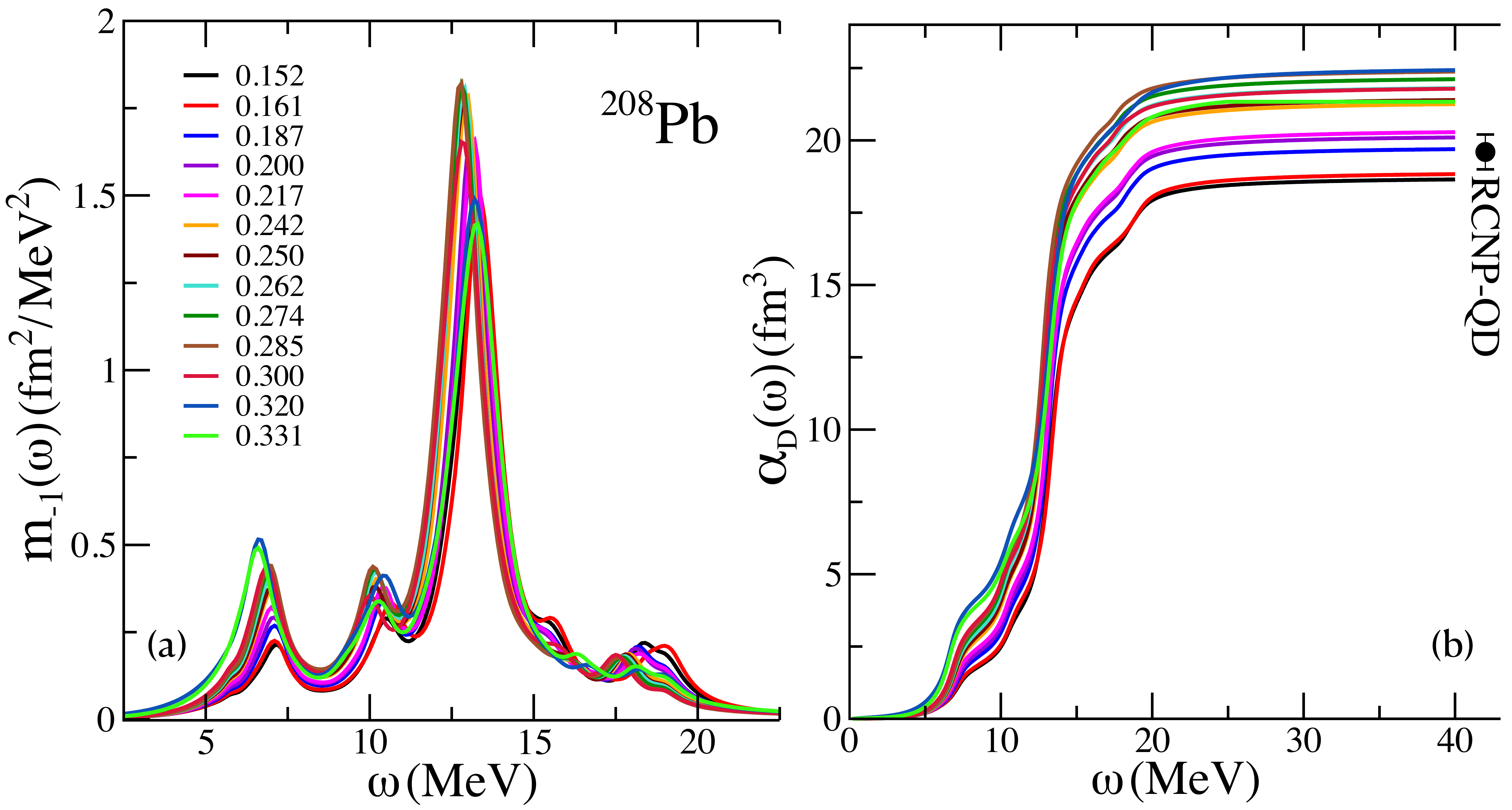}
 \caption{(Color online). (a) Inverse energy weighted dipole response of ${}^{208}$Pb computed with a diverse
 set of energy density functionals that span a wide range of value for the neutron skin thickness of ${}^{208}$Pb,
 as displayed by the labels in the figure. (b) Running (or cumulative) sum of the electric dipole polarizability, with
 the electric dipole polarizability defined as \AlphaD{}$\equiv$\AlphaD{}$(\omega_{\rm max})$.}
\label{Fig3}
\end{figure}

To elucidate the source of this discrepancy, we start by displaying in Fig.\,\ref{Fig3}(a) the distribution of electric 
dipole strength weighted by the inverse of the excitation energy: $m_{-1}(\omega)\!\equiv\!R(\omega;E1)/\omega$. 
The set of 13 covariant EDFs are all consistent with ground state properties of finite nuclei, yet they are flexible 
enough in that they span a wide range of values of \Rskin{208}, as indicated by the labels in the figure. Unlike the 
EWSR that is largely model independent\,\cite{Harakeh:2001}, the inverse energy weighted sum displays a significant 
model dependence. Indeed, models with a stiff symmetry energy generate larger values for both $m_{-1}$ and 
\AlphaD\,\cite{Piekarewicz:2010fa}. This fact is illustrated in Fig.\,\ref{Fig3}(b) that displays the ``running (or cumulative) 
sum" \AlphaD{}$(\omega)$. The predicted value for electric dipole polarizability is given by 
\AlphaD{}$\equiv$\AlphaD{}$(\omega_{\rm max})$, with $\omega_{\rm max}\!=\!40\,{\rm MeV}$. Also included in 
Fig.\,\ref{Fig3}(b) is the slightly refined value extracted from the RCNP experiment. The experimental distribution 
of strength contains a small, non-resonant ``contaminant"---the so-called quasi-deuteron (QD) contribution---at 
high energies that is not accounted for in the theoretical (RPA) predictions. Hence, for a meaningful comparison 
against experiment, the relatively small ($2.5\%$) quasi-deuteron contribution was removed, leading to a
revised estimate of \,\cite{Roca-Maza:2015eza}
\begin{equation}
  \text{\AlphaD{208}}\!=\!(19.6\pm0.6)\,{\rm fm}^{3}.
 \label{RCNPQD}
\end{equation}
This is the experimental value displayed in Fig.\,\ref{Fig3}(b).

\begin{figure}[ht]
 \centering
 \includegraphics[width=0.4\textwidth]{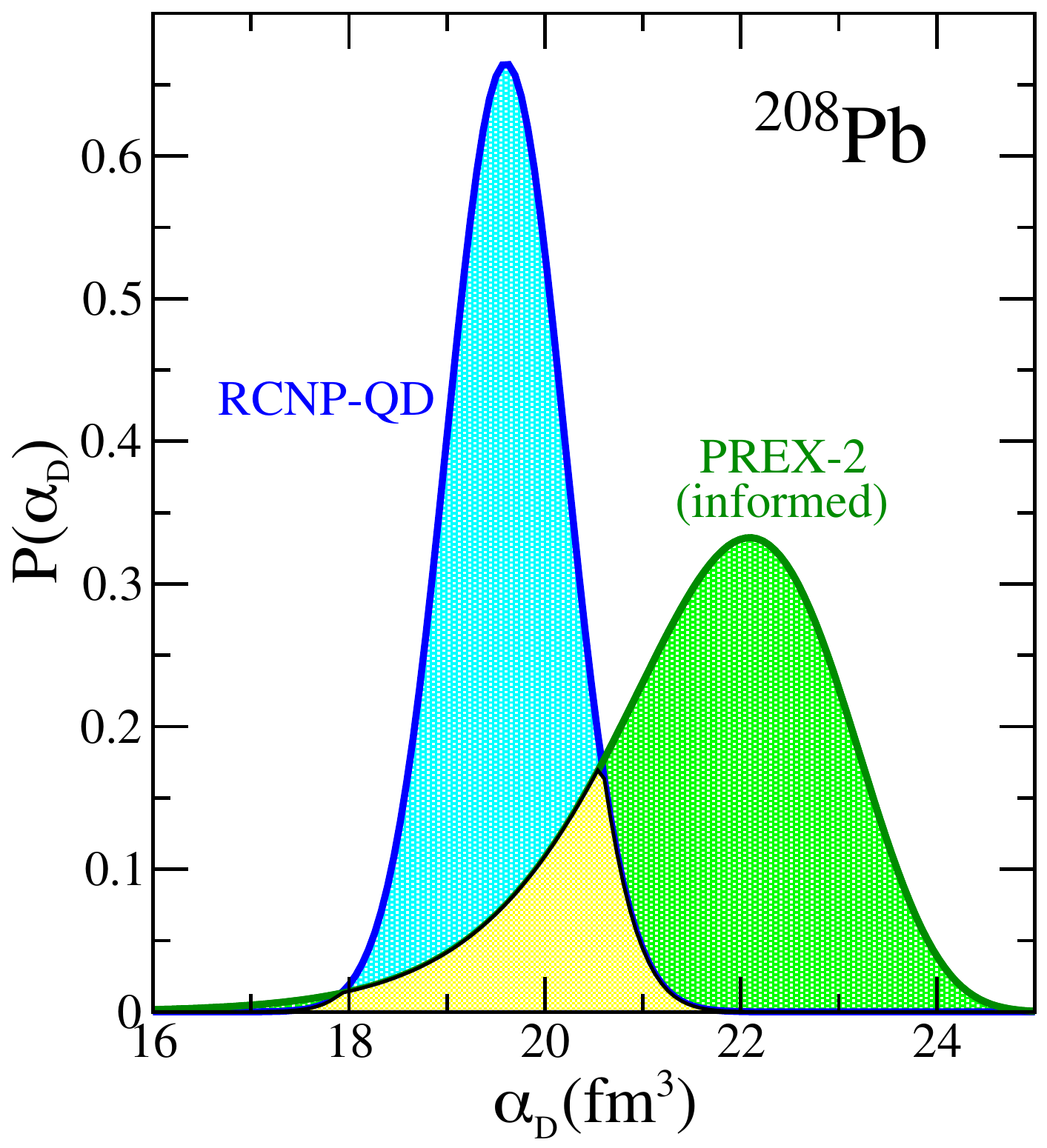}
 \caption{(Color online). The normal probability distribution as reported by the RCNP 
  experiment\,\cite{Tamii:2011pv}---minus the quasi-deuteron (QD) contribution\,\cite{Roca-Maza:2015eza}---is
  compared against the normal ratio distribution informed by the PREX-2 measurement\,\cite{Adhikari:2021phr}.
  The region shaded in yellow represents the area under the curve common to both probability distributions and
  amounts to less than 0.25 out of a possible maximum of 1.}
\label{Fig4}
\end{figure}

To properly quantify the discrepancy in the value of \AlphaD{208} extracted from the RCNP
measurement relative to the one informed by PREX-2, we display in Fig.\,\ref{Fig4} the associated 
probability distribution functions. For the RCNP experiment, we assume a normal (or Gaussian) 
probability distribution with the mean and standard deviation given in Eq.(\ref{RCNPQD}). Instead, 
for the PREX-2--informed result one obtains a ``normal ratio distribution"---similar in shape to the 
well-known skew-normal distribution. Following the derivation provided in the Appendix, one obtains 
the following PREX-2--informed estimate:
\begin{equation}
  \text{\AlphaD{208}}\!=\!(21.8^{+1.1}_{-1.4})\,{\rm fm}^{3},
 \label{AlphaDPrex2}
\end{equation}
where \AlphaD{208}$=\!21.8\,{\rm fm}^{3}$ is the median of the distribution, and the lower and
upper limits contain approximately 16\% and 84\% of the area under the curve. The figure 
illustrates the tension that emerges in comparing the two probability distributions. Indeed, the 
overlap region in Fig.\,\ref{Fig4}, estimated as the area under the curve shared by the two 
probability distributions, amounts to 0.25 out of a possible maximum of 1. An alternative definition
of the variation distance between two probability distributions is given in terms of the the largest 
possible difference at a given point. For Fig.\,\ref{Fig4}, the largest possible difference is obtained
at the peak of the RCNP-QD distribution, which results in a variation distance of 0.6. A perfect overlap 
between the two probability distribution would be equal to zero. In isolation, Fig.\,\ref{Fig4} may 
not be too much of a concern. However, it is the systematic discrepancy across various nuclei that 
we identify as the source of the tension.

\subsection{Electric Dipole Response of ${}^{48}$Ca}

\begin{figure}[ht]
 \centering
 \includegraphics[width=0.45\textwidth]{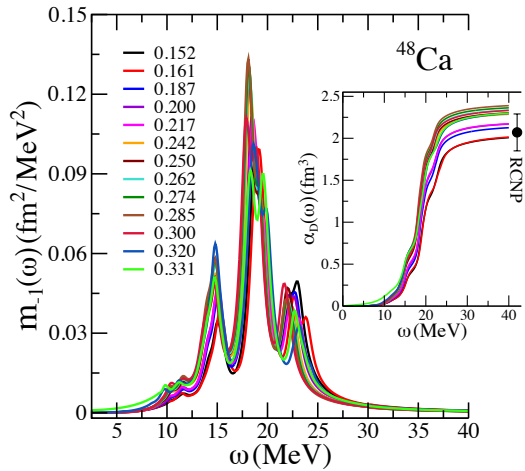}
\caption{(Color online). Inverse energy weighted dipole response of ${}^{48}$Ca computed with a diverse
 set of energy density functionals that span a wide range of value for the neutron skin thickness of ${}^{208}$Pb,
 as displayed by the labels in the figure. The inset displays the running (or cumulative) sum of the electric dipole 
 polarizability, with the electric dipole polarizability defined as \AlphaD{}$\equiv$\AlphaD{}$(\omega_{\rm max})$.}
\label{Fig5}
\end{figure}

\begin{figure}[ht]
 \centering
 \includegraphics[width=0.4\textwidth]{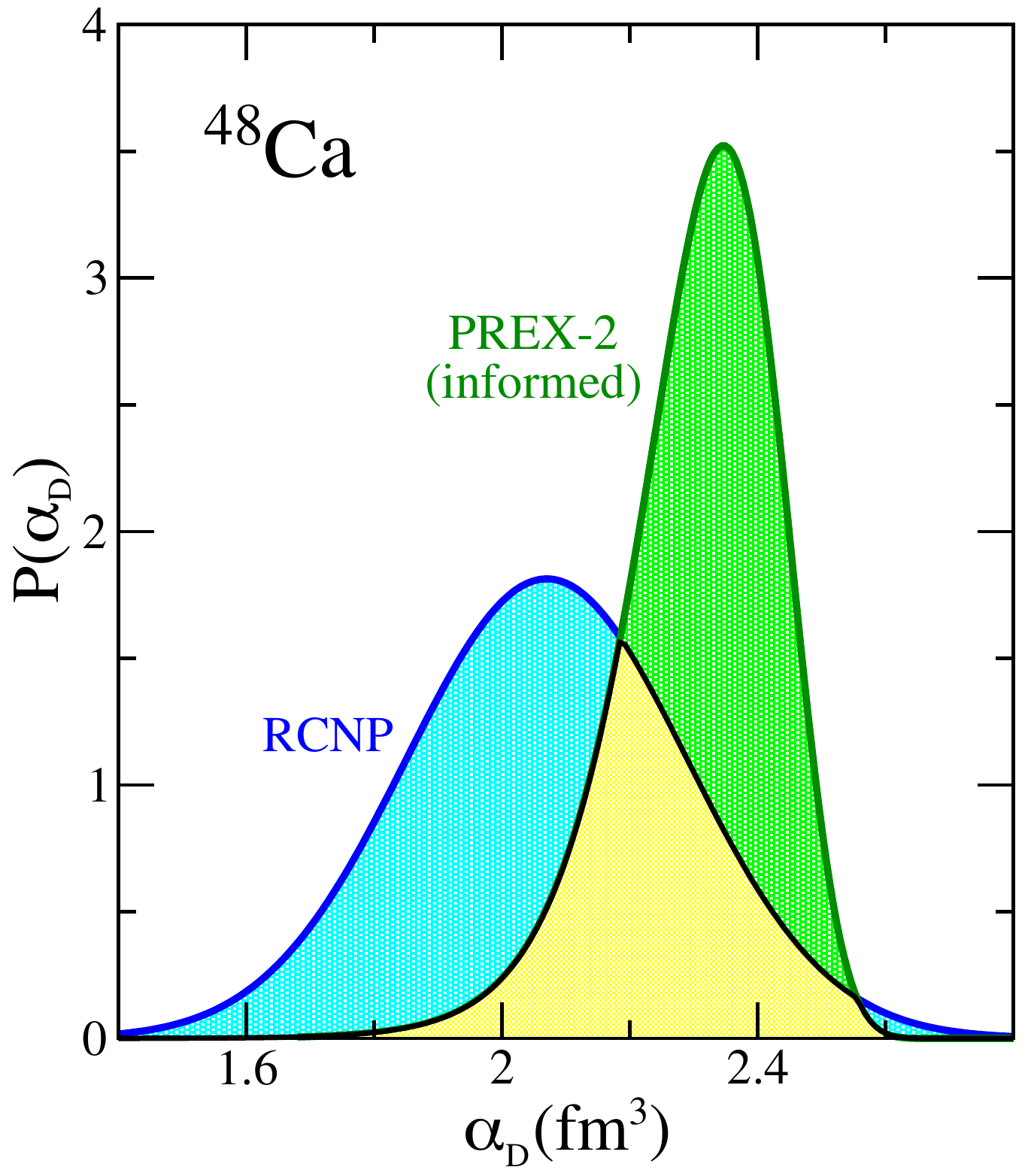}
 \caption{(Color online). The normal probability distribution as reported by the RCNP 
  experiment\,\cite{Birkhan:2016qkr} is compared against the normal ratio distribution 
  informed by the PREX-2 measurement\,\cite{Adhikari:2021phr}. The region shaded in 
  yellow represents the area under the curve common to both probability distributions 
  and amounts to about 0.5 out of a possible maximum of 1.}
\label{Fig6}
\end{figure}

As already mentioned, ${}^{48}$Ca provides a powerful bridge between microscopic approaches 
and density functional theory. At the beginning of the section we established that coupled-cluster 
predictions for the neutron skin of ${}^{48}$Ca underestimate the limits informed by PREX-2. In
this section we confront both sets of predictions, but now for the electric dipole 
polarizability\,\cite{Hagen:2015yea}. Following the same approach as in the case of ${}^{208}$Pb, 
the distribution of inverse-energy-weighted strength $m_{-1}(\omega)$ is displayed in Fig.\,\ref{Fig5}, 
alongside the running sum and the experimental value of \AlphaD{48} reported by Birkhan and 
collaborators\,\cite{Birkhan:2016qkr}. The labels in the figure continue to identify the various models
by the predictions for \Rskin{208} (not \Rskin{48}). The inset in the figure confirms that the experimental 
value of \AlphaD{48} favors models with a relatively soft symmetry energy, although the case is not as
convincing as for ${}^{208}$Pb because of the fairly large error bar. The impact of the large error bar
can be better appreciated in Fig.\,\ref{Fig6} where the overlap region between the two probability 
distributions has now increased to about 0.5.
Experimental values for \AlphaD{48} can now be collected from the RCNP experiment\,\cite{Birkhan:2016qkr}, 
the ab-initio/coupled-cluster predictions of Hagen and collaborators\,\cite{Hagen:2015yea}, and from the one 
obtained here by invoking the PREX-2 measurement. One obtains,
\begin{equation}
  \text{\AlphaD{48}}({\rm fm}^{3}) = 
   \begin{cases}
     2.07\pm0.22, & \text{RCNP\,\cite{Birkhan:2016qkr}}\\
     2.39\pm0.21, & \text{ab-initio\,\cite{Hagen:2015yea}} \\
     2.32^{+0.10}_{-0.13}, & \text{this work}.
   \end{cases}
   \label{AlphaD48}
\end{equation}

Although there is a noticeable difference in the central values between theory and experiment, the error 
bars are simply too large to draw any firm conclusion. This is an area in which the proposed, high-intensity
Gamma Factory at CERN could be of enormous value\,\cite{Placzek:2019xpw,Placzek:2020bjl,Piekarewicz:2021qpm}.

It is important to note that whereas the prediction from the ab-initio formalism for \AlphaD{48} is fully consistent 
with the one suggested in this work, the corresponding predictions for \Rskin{48} differ significantly, 
further suggesting that the strong isovector indicator is $J$\AlphaD{48} and not \AlphaD{48} alone. Indeed, the 
extracted value of $J\!\approx\!38\,{\rm MeV}$ from the PREX-2 measurement is considerably larger than the 
$J\!\lesssim\!30\,{\rm MeV}$ prediction from ab-initio calculations\,\cite{Hagen:2015yea}. 

\subsection{Electric Dipole Response of ${}^{68}$Ni}

The previous two sections were devoted to address the impact of the new PREX-2 measurement on 
the electric dipole response of ${}^{208}$Pb and ${}^{48}$Ca---the only two stable, neutron-rich, 
doubly-magic nuclei in the entire nuclear chart. These two nuclei provide valuable anchors for the 
calibration and interpretation of experiments on exotic nuclei with large neutron skins 
at future radioactive beam facilities\,\cite{Thiel:2019tkm}. Given the enormous challenges involved
in the commissioning and implementation of parity-violating experiments, photoabsorption reactions 
on exotic nuclei remain the only electroweak alternative to probe the density dependence of the
symmetry energy.

\begin{figure}[ht]
 \centering
 \includegraphics[width=0.45\textwidth]{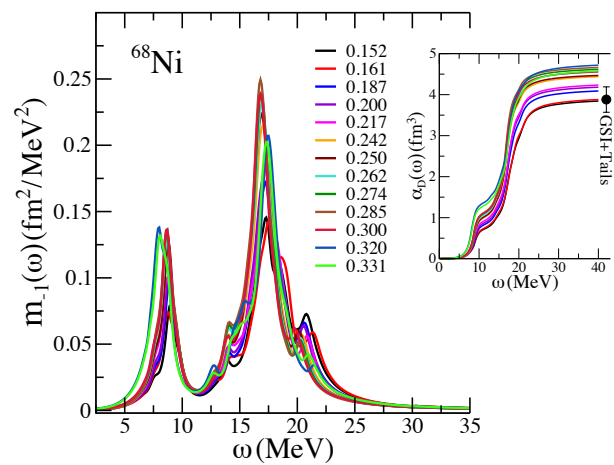}
\caption{(Color online). Inverse energy weighted dipole response of ${}^{68}$Ni computed with a diverse
 set of energy density functionals that span a wide range of value for the neutron skin thickness of ${}^{208}$Pb,
 as displayed by the labels in the figure. The inset displays the running (or cumulative) sum of the electric dipole 
 polarizability, with the electric dipole polarizability defined as \AlphaD{}$\equiv$\AlphaD{}$(\omega_{\rm max})$.}
\label{Fig7}
\end{figure}

The electric dipole response of the unstable, neutron-rich nucleus ${}^{68}$Ni has been measured 
using Coulomb excitation in inverse kinematics at the GSI facility in Germany\,\cite{Wieland:2009,Rossi:2013xha}. 
The electric dipole polarizability of ${}^{68}$Ni was reported to be\,\cite{Rossi:2013xha}:
\begin{equation}
  \text{\AlphaD{68}} = (3.40\pm0.23)\,{\rm fm}^{3} \xrightarrow[{\rm ``Tails"}]{}
  (3.88\pm0.31)\,{\rm fm}^{3}.
  \label{AlphaD68}
\end{equation}
The first number represents the experimental value obtained from 
integrating the photoabsorption cross section between 7.8 and 28.4 MeV. In turn, the 
second number accounts for the extrapolations to low- and high-energy that are required 
for a meaningful comparison between experiment and theory\,\cite{Roca-Maza:2015eza}. 
This second, slightly larger, value is denoted by ``GSI+Tails" on the inset of Fig.\,\ref{Fig7}. 

An interesting feature displayed in Fig.\,\ref{Fig7} is the significant amount of low-energy (``Pygmy") strength 
that is identified below the giant resonance region. Although the interpretation of its character is still under 
debate\,\cite{Paar:2007bk,Paar:2010ww,Carbone:2010az,Savran:2013bha,Bracco:2019gza}, the pygmy 
dipole resonance is often portrayed as an oscillation of the excess neutrons against the isospin symmetric 
core. Besides its interest as an exotic mode of excitation, the emergence of low-energy dipole strength 
has been found to correlate strongly with the development of a neutron rich skin\,\cite{Piekarewicz:2006ip}. 
This finding provides a compelling connection between the electric dipole polarizability and the neutron skin 
thickness---two critical observables used in the determination of the slope of the symmetry energy. Based on 
the results displayed in the inset of Fig.\,\ref{Fig7}, the electric dipole response of ${}^{68}$Ni continues 
to suggest a fairly soft symmetry energy. 

And not surprisingly, a tension develops between the results extracted directly from the GSI experiment and 
those constrained by PREX-2. In the case of the GSI experiment, the following value for the neutron skin 
thickness of ${}^{68}$Ni was reported\,\cite{Rossi:2013xha}:
\begin{equation}
 \text{\Rskin{68}} = (0.17\pm0.02)\,{\rm fm},
 \label{RskinGSI}
\end {equation} 
a value that is notably below the PREX-2--informed value of \Rskin{68}$\approx\!0.27\,{\rm fm}$ quoted 
in Eq.(\ref{Rskin68}).
\begin{figure}[ht]
 \centering
 \includegraphics[width=0.4\textwidth]{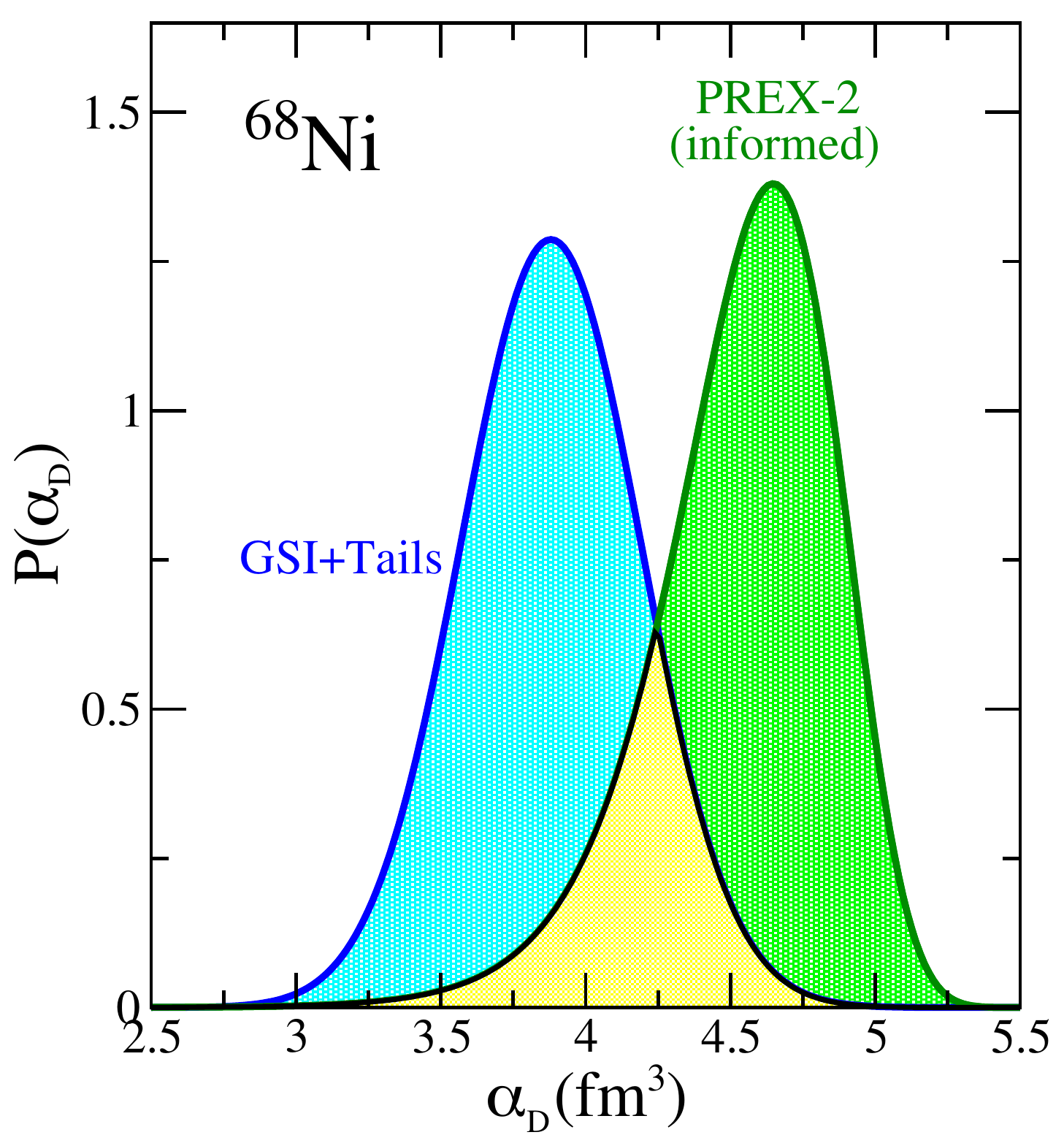}
 \caption{(Color online). The normal probability distribution as reported by the GSI  
  experiment\,\cite{Rossi:2013xha}---plus the addition of low- and high-energy tails\,\cite{Roca-Maza:2015eza}---is
  compared against the normal ratio distribution informed by the PREX-2 measurement\,\cite{Adhikari:2021phr}.
  The region shaded in yellow represents the area under the curve common to both probability distributions and
  amounts to less than 0.3 out of a possible maximum of 1.}
\label{Fig8}
\end{figure}
We conclude this subsection by displaying in Fig.\,\ref{Fig8} the experimental probability distribution alongside 
the normal ratio distribution inferred from PREX-2. Consistent with earlier findings, the latter is shifted to higher 
values relative to the former, with an overlap region that amounts to only 28\%. That is,
\begin{equation}
  \text{\AlphaD{68}}({\rm fm}^{3}) = 
   \begin{cases}
     3.88\pm0.31, & \text{GSI+Tails\,\cite{Rossi:2013xha,Roca-Maza:2015eza}}\\
     4.58^{+0.26}_{-0.33}, & \text{this work}.
   \end{cases}
   \label{AlphaD68}
\end{equation}

\subsection{Electric Dipole Response of ${}^{132}$Sn}

\begin{figure}[ht]
 \centering
 \includegraphics[width=0.45\textwidth]{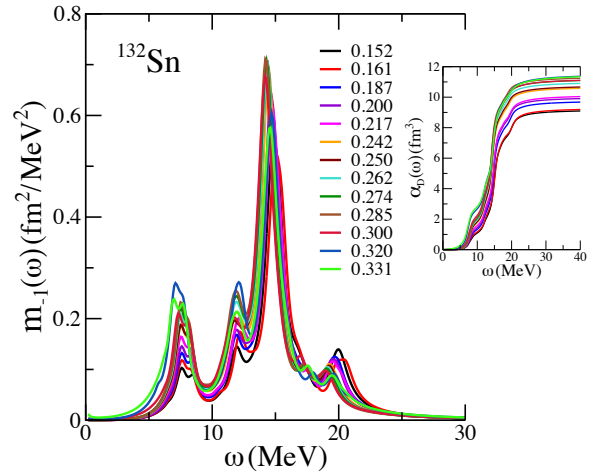}
\caption{(Color online). Inverse energy weighted dipole response of ${}^{132}$Sn computed with a diverse
 set of energy density functionals that span a wide range of value for the neutron skin thickness of ${}^{208}$Pb,
 as displayed by the labels in the figure. The inset displays the running (or cumulative) sum of the electric dipole 
 polarizability, with the electric dipole polarizability defined as \AlphaD{}$\equiv$\AlphaD{}$(\omega_{\rm max})$.}
\label{Fig9}
\end{figure}

Lastly, predictions are displayed in Fig.\,\ref{Fig9} for the unstable, doubly-magic, neutron-rich nucleus ${}^{132}$Sn. 
Although a measurement of the electric dipole response of ${}^{132}$Sn---both in the low- (Pygmy) and high-energy 
(Giant) regions---has been reported in Ref.\,\cite{Adrich:2005}, a precise value of the electric dipole polarizability is 
not yet available. Nevertheless, given that several experimental campaigns are being planned to measure the 
electric dipole response of ${}^{132}$Sn, it is pertinent to provide some predictions. Based on the dipole response 
of ${}^{68}$Ni, one would expect the emergence of a significant amount of low-energy strength. In fact, it was shown in 
Ref.\,\cite{Piekarewicz:2006ip} how the emergence of low-energy dipole strength correlates with the development 
of a neutron-rich skin along the tin isotopes, from ${}^{100}$Sn all the way to ${}^{132}$Sn. However, it was also 
observed that as the $h_{11/2}$ neutron orbital was progressively filled, the low-energy strength ``saturates" at 
${}^{120}$Sn, indicating that high-angular momentum orbitals play a minor role in driving low-energy transitions 
of low multipolarity (see Figs.\,4-5 of Ref.\,\cite{Piekarewicz:2006ip}). One of the main virtues of ${}^{132}$Sn relative 
to the lighter open-shell isotopes lies in its simple structure since, as a doubly-magic nucleus, ${}^{132}$Sn is 
insensitive to the (presently unclear) role of pairing correlations\,\cite{Bassauer:2020iwp}.

A precise measurement of the electric dipole polarizability of ${}^{132}$Sn is well motivated given the
two model independent correlations connecting ${}^{132}$Sn to ${}^{208}$Pb identified in 
Ref.\cite{Piekarewicz:2012pp}, one for the neutron skin thickness and the other one for the electric 
dipole polarizability. This would suggest that a measurement of \Rskin{132} (if feasible) should mirror the 
PREX-2 result and yield a correspondingly large value for the neutron skin thickness of ${}^{132}$Sn,
confirming that the symmetry energy is stiff. On the other hand, if one follows the \AlphaD{} correlation, 
then the relatively low value of \AlphaD{208} reported by the RCNP experiment would also imply a low 
value for  electric dipole polarizability of ${}^{132}$Sn, suggesting instead that the symmetry energy is soft.

The above situation can be precisely quantified. For example, invoking the RCNP measurement of \AlphaD{208}, 
one would infer on the basis of the strong correlations uncovered in Ref.\,\cite{Piekarewicz:2012pp} the following values 
for ${}^{132}$Sn:
\begin{subequations}
\begin{align}
 & \text{\Rskin{132}}\!=\!(0.23\pm0.02)\,{\rm fm}, \\
 & \text{\AlphaD{132}}\!=\!(10.08\pm0.15)\,{\rm fm}^{3}.
\end{align}
\end{subequations}
In contrast, if one relies on the recent PREX-2 analysis, then one infers a normal ratio distribution from which the 
significantly larger values are obtained:
\begin{subequations}
\begin{align}
 & \text{\Rskin{132}}\!=\!(0.35\pm0.07)\,{\rm fm}, \\
 & \text{\AlphaD{132}}\!=\!(11.07^{\,+0.75}_{-0.97})\,{\rm fm}^{3}.
\end{align}
\end{subequations}

\section{Conclusions}
\label{Sec:Conclusions}

Borrowing a term from cosmology, the density ladder is a succession of theoretical, experimental,
and observational techniques aimed to determine the equation of state of neutron star matter at 
increasingly higher densities. Laboratory experiments sensitive to the dynamics of neutron rich 
matter in the vicinity of nuclear saturation density provide the first rung of such a ladder. In this
context, electroweak probes provide the cleanest connection to the equation of state. 

Earlier estimates of the symmetry energy based on measurements of the electric dipole polarizability 
suggest that the symmetry energy is relatively soft. The softness of the symmetry energy was later 
validated by theoretical approaches as well as a large suite of experiments\,\cite{Lattimer:2012nd,Drischler:2020hwi}. 
However, the suggestion of a soft symmetry energy has now been brought into question by the recent 
report of an unexpectedly large neutron skin thickness in ${}^{208}$Pb\,\cite{Adhikari:2021phr}. By 
exploiting the strong correlation between \Rskin{208} and $L$, it has been suggested that the symmetry 
energy is, instead, fairly stiff\,\cite{Reed:2021nqk}. Connecting PREX-2 to various neutron stars properties 
was recently explored in Refs.\,\cite{Reed:2021nqk,Essick:2021kjb}.

In this paper we assessed the impact of PREX-2 on the electric dipole polarizability 
of several closed-shell nuclei and compared our results against those values extracted directly from experiment. 
A set of covariant energy density functionals that span a wide range of values of \Rskin{208} was used to compute 
ground state densities and the electric dipole response of ${}^{208}$Pb, ${}^{48}$Ca, ${}^{68}$Ni, and 
${}^{132}$Sn. Relying on the nearly perfect correlation between \Rskin{} and both 
$J$\AlphaD{}\,\cite{Roca-Maza:2013mla} and $J$\,\cite{Reed:2021nqk} for the class of models explored
in this work, we derived limits on \AlphaD{}  informed by PREX-2. In all instances, the electric dipole polarizability
informed by PREX-2 overestimated the corresponding values obtained directly from earlier measurements 
of the dipole response. So while direct measurements of the electric dipole polarizability seem to suggest that 
the symmetry energy is soft at nuclear densities, the PREX-2--informed values suggest the opposite. Given 
the vital role that terrestrial experiments play in constraining the slope of the symmetry energy $L$, the 
resolution of this dilemma is of utmost importance. 

Naturally, one would like to see a significant reduction in the experimental uncertainty. However, 
prospects of a more precise electroweak determination of \Rskin{208} are slim, given that the PREX 
campaign is now over. A factor of two improvement may be possible at the Mainz Energy-recovery 
Superconducting Accelerator (MESA), but the timing is not optimal as the facility is currently under 
construction\,\cite{Becker:2018ggl}. The prospects for improving the precision in the determination 
of \AlphaD{} are significantly better with the commissioning of a Gamma Factory at CERN, the 
implementation of reactions with relativistic radioactive beams (R3B) at FAIR, and the combination 
of high yields and high energies for very neutron-rich isotopes that will be available at FRIB400. The 
challenges in this arena are associated with the production of unstable nuclei with large skins as well 
as the determination of the electric dipole response over a wide range of energies. In particular, the 
low-energy region---above and below the neutron separation energy---is of critical importance given 
the significant contribution of the soft dipole (Pygmy) mode to the electric dipole polarizability. As we
enter the golden age of neutron-star physics\,\cite{Mann:2020}, a sustained and concerted community 
effort in both experiment and theory is both timely and important.

\appendix*\section{}
The aim of this appendix is to derive the probability distribution $f(x)$ for the ratio of two normally 
distributed observables ($x_{1}$ and $x_{2}$) that are perfectly correlated. The resulting distribution 
$f(x)$ displays a ``normal ratio distribution", which in the case of Figs.\,\ref{Fig4}, \ref{Fig6}, and \ref{Fig8}
strongly resembles the well-known skew-normal probability distribution.

The derivation of $f(x)$ hinges on the strong correlation---at least within the models employed in this 
work---between the neutron skin thickness of ${}^{208}$Pb and both the symmetry energy at saturation 
density $J$\,\cite{Reed:2021nqk} and the product of $J$ times the electric dipole polarizability displayed 
in Fig.\,\ref{Fig2}. Conceptually, computing the probability distribution is easy to understand from the 
perspective of statistical sampling. Assuming a Gaussian distribution for the neutron skin thickness of 
${}^{208}$Pb with a mean of $\mu\!=\!0.283\,{\rm fm}$ and a standard deviation of 
$\sigma\!=\!0.071\,{\rm fm}$\,\cite{Adhikari:2021phr}, one draws individual samples of \Rskin{208} using, 
for example, the Box--Muller transform. Then, assuming that the correlation between \Rskin{208} and both 
$J$ and $J$\AlphaD{} is perfect (as they nearly are!) one can in turn generate the associated values for $J$ 
and $J$\AlphaD{}. Then, by simply dividing the latter over the former one obtains individual values for \AlphaD{}
that obey a normal ratio distribution as displayed in Figs.\,\ref{Fig4}, \ref{Fig6}, and \ref{Fig8}. 

However, a closed-form expression may also be derived for the normal ratio distribution.
To do so, one starts with a standard Gaussian distribution of zero mean and unit standard 
deviation. That is,
\begin{equation}
  f(\epsilon)=\frac{1}{\sqrt{2\pi}}\,{\mathlarger e}^{-\epsilon^{2}/2}.
 \label{fnormal}
\end {equation} 
Any normal distribution $f(x)$ with a mean of $\mu$ and standard deviation of $\sigma$ can be
readily mapped into the above distribution by simply letting $x\!=\!\mu+\sigma\epsilon$. Let us 
then assume the existence of two perfectly correlated observables $X_{1}$ and $X_{2}$ normally
distributed with means $\mu_{1}$ and $\mu_{2}$ and standard deviations $\sigma_{1}$ and 
$\sigma_{2}$. For a perfect correlation between observables, one can write their ratio as follows:
\begin{equation}
  \frac{x_{2}}{x_{1}} \equiv x = \left(\frac{\mu_{2}+\sigma_{2}\epsilon}{\mu_{1}+\sigma_{1}\epsilon}\right)
  \Rightarrow \epsilon = - \left(\frac{\mu_{2}-\mu_{1}x}{\sigma_{2}-\sigma_{1}x}\right).
 \label{NRD}
\end {equation} 
Note that the perfect correlation between $X_{1}$ and $X_{2}$ is encoded in the following
relations:
\begin{subequations}
\begin{align}
  & \frac{x_{2}-\mu_{2}}{\sigma_{2}} = \epsilon = \frac{x_{1}-\mu_{1}}{\sigma_{1}} \Leftrightarrow \\
  &  x_{2}=\left(\mu_{2}-\frac{\sigma_{2}}{\sigma_{1}}\mu_{1}\right)
                              +\left(\frac{\sigma_{2}}{\sigma_{1}}\right)x_{1}.
 \label{x2x1correlation}
\end{align}
\end{subequations} 
That is, the statistical sampling of $f(\epsilon)$ generates values of $\epsilon$ that determine $x_{1}$ 
which, in turn, determines $x_{2}$---through the linear correlation. One can finalize the derivation of
the normal ratio distribution using Eq.(\ref{NRD}). That is,
\begin{equation}
f(\epsilon)d\epsilon = f(\epsilon(x))\Big|\frac{d\epsilon}{dx}\Big| dx = f(x)dx, 
\end{equation}
from where the normal ratio distribution is obtained:
\begin{equation}
 f(x)\!=\!\frac{1}{\sqrt{2\pi}} \frac{|\mu_{1}\sigma_{2}\!-\!\mu_{2}\sigma_{1}|}{(\sigma_{2}\!-\!\sigma_{1}x)^{2}}
 \exp\!\left[\!-\frac{1}{2}\!\left(\frac{\mu_{2}\!-\!\mu_{1}x}{\sigma_{2}\!-\!\sigma_{1}x}\right)^{\!2}\right].
\end{equation}
\vfill

\begin{acknowledgments}
 I thank Prof. Antonio Linero for guiding me through the analytic derivation of the normal ratio 
 distribution. I also thank Dr. Pablo Giuliani for many useful comments and a critical reading of
 the manuscript. This material is based upon work supported by the U.S. Department of Energy 
 Office of Science, Office of Nuclear Physics under Award DE-FG02-92ER40750. 
\end{acknowledgments} 

\bibliography{PREXI2on AlphaD.bbl}
\end{document}